\documentclass[reprint,nofootinbib,preprintnumbers,twocolumn,floatfix,superscriptaddress,prl]{revtex4-2}
\usepackage{physics}
\usepackage[clock]{ifsym}
\usepackage{natbib}
\usepackage{amsmath,amssymb}
\usepackage{graphicx}
\usepackage{adjustbox}
\usepackage[dvipsnames]{xcolor}
\usepackage[super]{nth}
\usepackage[caption=false]{subfig}
\usepackage{float}
\usepackage[pdftex]{hyperref} 
\setcitestyle{square,numbers}
\usepackage{dsfont}

\begin{document}

 \title{Probing classical and quantum violations of the equivalence of active and passive gravitational mass}
	\author{Vasileios Fragkos}
    \email[]{vasileios.fragkos@fysik.su.se}
	\affiliation{Department of Physics, Stockholm University, SE-106 91 Stockholm, Sweden}
	\author{Igor Pikovski}
    \email[]{igor.pikovski@fysik.su.se}
	\affiliation{Department of Physics, Stockholm University, SE-106 91 Stockholm, Sweden}
	\affiliation{Department of Physics, Stevens Institute of Technology, Hoboken, NJ 07030, USA}
	\date{\today}

\begin{abstract}
The equivalence of active and passive (EAP) gravitational mass is one of the most fundamental principles of gravity. But in contrast to the usual equivalence of inertial and (passive) gravitational mass, the EAP has not received much attention. 
Here we revisit this principle and show how it can be used to probe quantum gravity in laboratory-based experiments. We first examine how the dynamics under EAP violations affects classical systems and show that new laboratory tests can be performed, to improve over the current experimental bounds and to test new manifestations of EAP violations. We then extend the analysis to the quantum domain, where quantized energy contributes to mass and the EAP principle can thus shed light on how quantum source masses would gravitate. We show that experiments with cold polar molecules, and future experiments with nuclear atomic clocks, can test the quantum EAP in a regime where quantum gravity phenomenology could become relevant. 
Our results open new opportunities for fundamental tests of gravity in high-precision laboratory experiments that can shed light on foundational principles of gravity and its interface with quantum theory.
\end{abstract}

 \maketitle

\section{Introduction}
The equivalence principle is one of the cornerstones of general relativity and has been the focus of many experiments  \cite{Wagner_2012tests.EP,MicroscopeEP,kasevich.Atom.interf.for.EP,tino2020precision}. Yet, another fundamental property of mass is of similar significance, but is far less explored: the equivalence between active and passive gravitational mass (EAP), $m_a = m_p$. It ensures that the strength of \textit{sourcing} gravity is exactly the same as the strength of \textit{feeling} gravity by a probe system. This exact equivalence is central to modern physics \cite{bondi1957negative,will1976active,treder2013fundamental,Muller.A-P.electric.charge}, underpinning both gravity and Newton's third law of action and reaction \cite{treder1983galilei,ohanian2010energy}. But the only dedicated laboratory experiment probing this principle was performed by Kreuzer in 1968, constraining the relative EAP violation parameter
\begin{equation} \label{eq:S}
    S(1,2)= \frac{m_{1a}}{m_{1p}} - \frac{m_{2a}}{m_{2p}}
\end{equation}
to $S(F, Br) \lesssim 5 \times 10^{-5}$ for fluorine and bromine, using a Cavendish-type setup with solid Teflon submerged in a fluid \cite{Kreuzer}. Lunar laser ranging (LLR) is the only other test of the EAP based on celestial dynamics of the moon \cite{Bartlett}, constraining $S(Fe,Al) \lesssim 4 \times 10^{-14}$ \cite{Singh} according to the moon's expected composition \cite{Moon.model.asymmetry.1972apollo, muller1972moon, 2-component-Moon.Wood1973}.

\begin{figure*}[t]
 \centering
 \includegraphics[width=.9\textwidth]{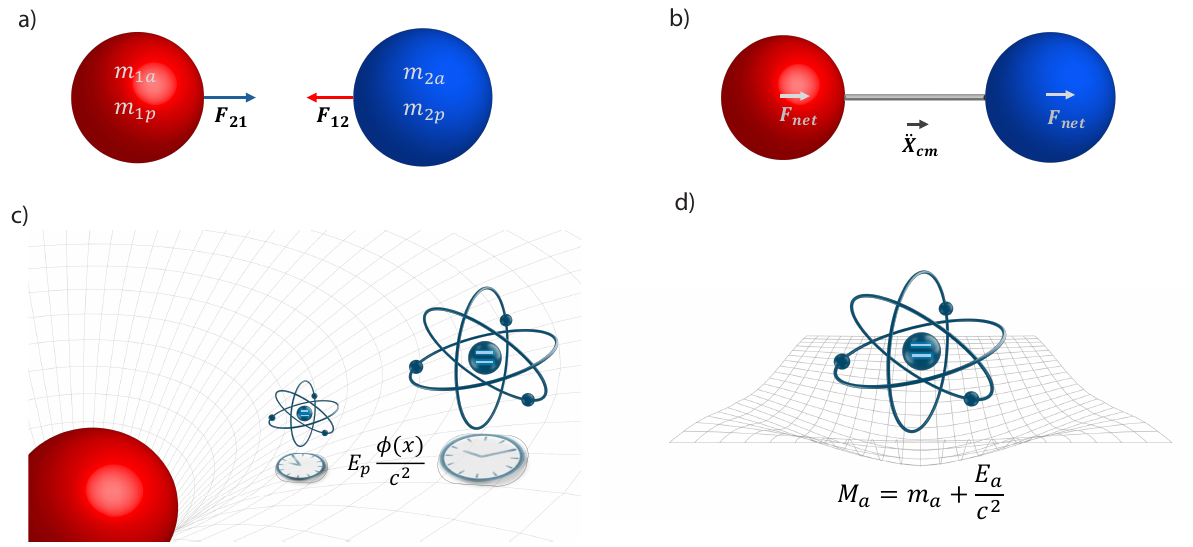}
 \caption{Illustration of violations of the equivalence of active and passive mass (EAP), and its manifestations in Newtonian, relativistic and quantum physics. a) Two objects exert gravitational forces on each other, governed by their active and passive masses. If EAP is violated, these mutual forces are not equal in magnitude. b) A composite system held by non-gravitational forces will experience a net force from EAP violations, resulting in a net self-acceleration. c) In general relativity, energies also contribute to the mass. This is observed directly through the redshift of clocks\cite{chou2010optical, bothwell2022resolving, zheng2023lab}, which results from the coupling of the clocks' passive energy to the external gravitational field. d) Probe systems themselves create a minute gravitational field, but too small to be observed directly. If the EAP is valid, the same gravitational charge that couples to the external field would also generate the gravitational field. In contrast, any discrepancy of the active and passive mass, including the quantized energy of the system, would require an EAP violation. This opens the window to probe how gravitational fields are sourced by quantized energy, by testing for a self-acceleration in AMO systems with internal superpositions.  
 }
 \label{Fig:summary}
\end{figure*}

The EAP is not directly related to the usual equivalence principle, which only states that inertial mass must equal the passive gravitational mass\footnote{Here we assume the exact validity of this regular equivalence principle.}, $m_i = m_p$. There are several reasons why testing the EAP is of importance. Not only is it an independent foundational principle of gravity, it also connects to unique features of both general relativity and its interface with quantum theory. In general relativity, the EAP is not explicit as the notion of passive mass loses its clear meaning from the Newtonian theory \cite{rindler2006relativity}. The gravitational ``charge'' is replaced by the stress-energy tensor $T_{\mu  \nu}$, and even localized gravitational energy. Tolman, for example, showed that the effective inertial mass in a static space-time takes the form $m_i= J_0$ \cite{tolman1934relativity} for the current $ J_{\mu} = -\int d^3x \sqrt{-g}\left(T^0_{\, \, \mu} + t^0_{\, \, \mu} \right)$, where $t^{\nu}_{\, \, \mu}$ is the Einstein pseudo-tensor that captures gravitational energy and momentum. But even though both energy and pressure contribute to the coupling to gravity and thus to mass, the EAP is expected to hold in general relativity such that $m_i=m_p=m_a$ remains valid. Thus the charge governing the source of gravity and the coupling to an external gravitational field should be the same -- which can be interpreted as a manifestation of Mach's principle in Einstein's theory \cite{treder2013fundamental}. Nevertheless, explicit calculations within general relativity led to seemingly anomalous contributions to active mass by different forms of energy, such as outlined by Tolman for when radiation pressure becomes relevant \cite{tolman1934relativity}. This was later resolved by Misner and Putnam \cite{MisnerPutnan}. 
Similarly, contradictory results on the explicit EAP manifestation or its violations in general relativity were obtained by Bonnor \cite{bonnor1992active} and Rosen and Cooperstock\cite{rosen1992mass}, and others \cite{nieuwenhuizen2007nonequivalence,ohanian2010energy,vollick2022meaning}. In alternative theories of gravity, such as Brans-Dicke theory, the EAP may well be violated \cite{askari1995mass,will2014confrontation}, as well as in some models of quantum gravity \cite{garfinkle1991charged}. EAP violations might also be linked to avoiding singularities \cite{von1979equivalence}. Even the  effective passive mass $m_p$ alone is a problematic concept in general relativity\cite{ohanian2010energy}: it is obtained perturbatively from the response of a system to external gravitational fields, but for composite systems with interactions and kinetic energies another seeming anomaly involving energies (which only cancel by the virial theorem) obtained by Eddington \cite{eddington1938problem}, Nordtvedt \cite{nordtvedt1970gravitational,nordtvedt1974equation} and others \cite{fischbach1981general, lebed2013gravitational} was only recently resolved \cite{carlip1998kinetic, PikovskiGravitationalMass}. The EAP $m_p = m_a$ in general relativity is thus subtle, especially when energies are involved, and possible violations could arise that can depend on nuclear structure or internal energies. Experimental tests of the EAP in different settings and composite systems can thus probe a critical pillar of general relativity that can also shed light on its possible alternatives.

The quantum nature of energy, and thus mass, is an additional important motivator for testing the EAP. As energy contributes to gravitational mass in general relativity, superpositions of masses\footnote{The superselection rule \cite{bargmann1954unitary,giulini1996galilei}, which is exact for electric charge, is not valid for mass beyond Galilean dynamics \cite{greenberger2001inadequacy,zych2019puzzling}. Superpositions of mass have to be taken into account when the weight of superposed energy according to $E=mc^2$ becomes relevant.} open new opportunities for fundamental tests of the interplay of quantum theory and gravity \cite{zych2011quantum,pikovski2015universal,pikovski2017time,zych2018.QUANTUM.EP,rosi2017quantum}, where the quantum systems act as passive probes. There has also been a surge of interest in studying \textit{quantum sources} of gravity, with the possibility of laboratory tests of quantum gravity \cite{schmole2016micromechanical,GIEBose,GIEMarlettoVedral}. However, while quantum probes of gravity are routinely used in experiments, such as with neutron matter-waves \cite{colella1975observation,sponar2021tests}, atomic fountains \cite{Muller2010APM,Tino:2020dsl,asenbaum2020atom} or atomic clocks \cite{chou2010optical, bothwell2022resolving, zheng2023lab}, quantum source masses remain far outside the reach of current experiments. But any distinction between the two must be accompanied with a breakdown of the symmetry between sources and probes -- akin to EAP violations. The goal to achieve experimental signatures of quantum sources is thus unnecessary \cite{overstreet2023inference}, unless some inequivalence between sourcing and probing gravity is postulated. This motivates the search for possible deviations between probing and sourcing gravity, such as the EAP principle, when quantum systems are involved. A direct test of quantum sources of gravity would amount to a direct test of quantum gravity, but EAP violations provide an alternative indirect test.  Indeed, in some models of quantum gravity phenomenology a breakdown of this principle and the associated validity of Newton's third law can occur \cite{borzeszkowski2012meaning}, such as due to quantum foam\cite{Bekenstein_2014} or within string theory\cite{garfinkle1991charged}. 

Here, we study how violations of active and passive gravitational mass can manifest themselves in the dynamics of interacting systems in both the classical and quantum domains, illustrated in figure \ref{Fig:summary}. We isolate the two distinct signatures, namely the inequality of the active and passive mass parameters, and the breakdown of Newton's third law, and how they can be probed in laboratory settings with unprecedented sensitivity. Our work highlights that new tests are possible, and that violations are not captured by the single parameter $S$ in eq. \eqref{eq:S}, especially for quantum systems. In analogy to possible purely quantum mechanical violations of the usual equivalence principle \cite{rosi2017quantum,zych2018.QUANTUM.EP}, we outline tests of the EAP involving quantum superpositions of energy which can shed light on the interplay of quantum theory and general relativity.

\begin{figure*}[t]
 \centering
 \includegraphics[width=.95\textwidth]{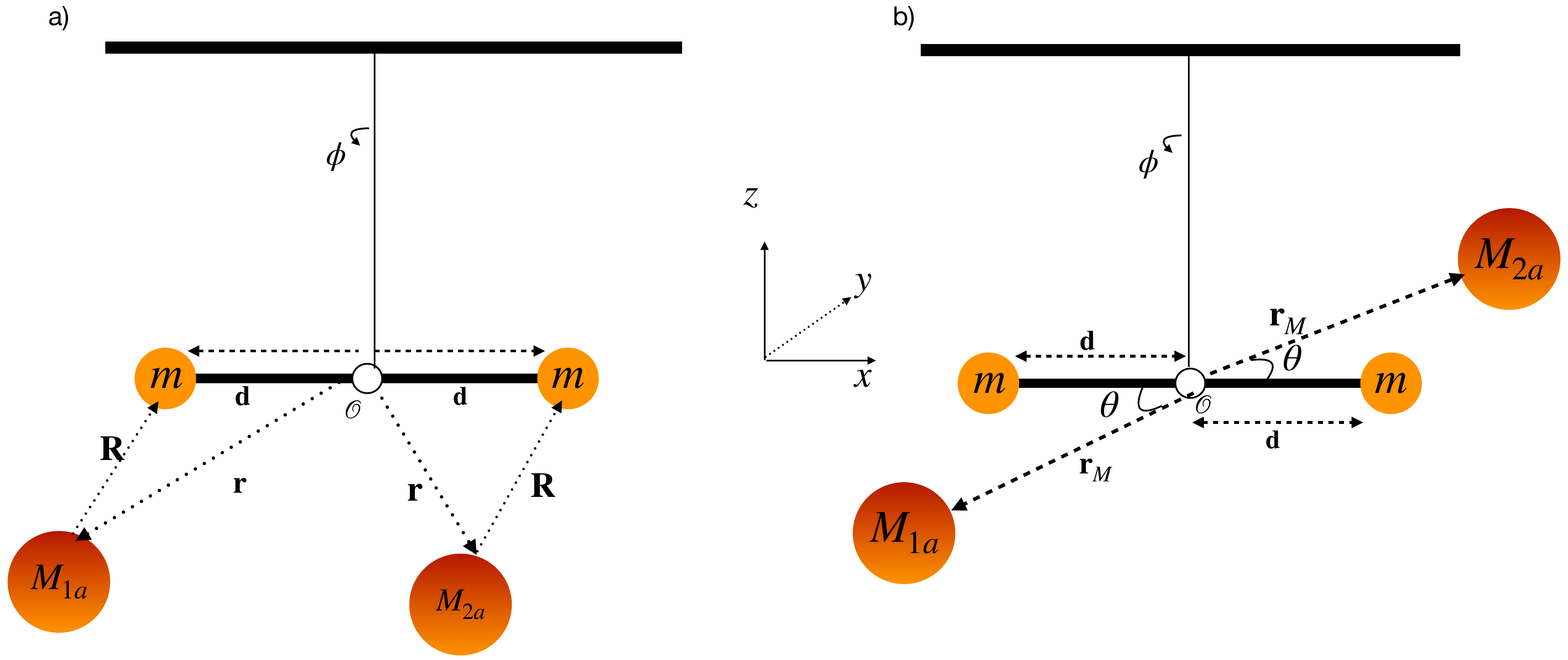}
 \caption{Proposed Cavendish-type experiments to constrain the two EAP violation parameters $S$ as first tested by Kreuzer \cite{Kreuzer} and $\sigma$ introduced here. $M_{1a}$ and $M_{2a}$ should be made of different materials to determine the material-dependent $S(1,2)$ and $\sigma(1,2)$.  The test masses $m$ are placed at equal distances $\boldsymbol{d}$ from the point $\mathcal{O}$ where the wire is attached, and their angular deflection acceleration $\ddot{\phi}$ is observed.
 The setup $(a)$ on the left results in a net-torque only if $S(1,2) \neq 0$ and does not depend on $\sigma(1,2)$. The configuration $(b)$ on the right is the standard Cavendish setup, but with different materials for the two big masses, which depends only on the parameter $\sigma(1,2)$. The mathematical details are given in Appendices B and C. 
 }
 \label{Cavendish}
\end{figure*}

\section{Dynamics under EAP violations}
We start by describing the dynamics with EAP violations. Principles of regular physics have to drastically change in order to accommodate such violations. For example, the usual Lagrangian or Hamiltonian methods cannot be applied, unless the laws to generate the equations of motion are also modified \cite{clausius1877potentialfunction}. This is because the use of potential functions implicitly assumes a symmetry between the interacting systems. Instead, we focus on EAP violations directly on the equations of motion. For two gravitating systems we have
\begin{equation} \label{eq:forces}
\begin{split}
    \boldsymbol{F}_{21}  & =Gm_{1p}m_{2a}\frac{\boldsymbol{x}_1-\boldsymbol{x}_2}{|\boldsymbol{x}_1-\boldsymbol{x}_2|^3}  \\
   \boldsymbol{F}_{12}  & = -Gm_{1a}m_{2p}\frac{\boldsymbol{x}_1-\boldsymbol{x}_2}{|\boldsymbol{x}_1-\boldsymbol{x}_2|^3} \, ,
\end{split}
\end{equation}
where in $\boldsymbol{F}_{ij}$ the first index refers to the system that sources the field, and the second index refers to the probe system reacting to this field.
To better see the result of this asymmetry, we can write the equations in terms of the usual Newtonian force with respect to the passive gravitational mass, $\boldsymbol{F}_N=-Gm_{1p} m_{2p} \frac{\boldsymbol{x}_1-\boldsymbol{x}_2}{|\boldsymbol{x}_1-\boldsymbol{x}_2|^3}$:
\begin{equation} \label{eq:forcesG}
\begin{split}
    \boldsymbol{F}_{21}  & = -\frac{\tilde{G}_2}{G} \boldsymbol{F}_N \\
    \boldsymbol{F}_{12}  & = \frac{\tilde{G}_1}{G} \boldsymbol{F}_N
\end{split}   
\end{equation}
Here we have introduced the material-dependent gravitational interaction strength $\tilde{G}_i \equiv G m_{i a} /m_{i p} $. The above expression shows that one can think of an EAP violation also in terms of a varying gravitational constant $G$ that depends on the material (a universal violation would simply be absorbed in the universal gravitational constant). This motivates Cavendish-type tests that probe the strength of gravity between different materials to probe the EAP, as first performed by Kreuzer \cite{Kreuzer}. It essentially amounts to determining independently the passive mass $m_p$, and then to measure the gravitational force exerted by this mass. A modern variation of this method is outlined in Figure \ref{Cavendish} . One can construct a Cavendish-type experiment with different source masses, but which have the exact same passive mass. This ensures a vanishing net torque on a probe system in the setup according to regular physics. But if the EAP is violated, a non-zero torque would arise. Table \ref{table:modified_cavendish} shows that such a setup with current capabilities can surpass the Kreuzer bounds in laboratory experiments and explore the EAP for various materials.

\begin{table}[b]
\centering
\begin{tabular}{ |p{3.3cm}||p{2.2cm}|p{2.2cm}|  }
 \hline
 \multicolumn{3}{|c|}{Proposed Modified Cavendish test of EAP} \\
 \hline
\hspace{0.5cm}Variables & \hspace{0.5cm} Case i & \hspace{0.6cm} Case ii\\
 \hline
 $|\ddot{\boldsymbol{\phi}}|\pm \Delta|\ddot{\boldsymbol{\phi}}|$  $[nrad \cdot s^{-2}]$ & $<0\pm 0.1$    &$<0\pm 0.001$ \\
 $d\pm \Delta d$ \hspace{0.4cm} $[cm]$&   $10\pm 0.001$  & $0.5\pm 0.01$   \\
$R\pm \Delta R$\hspace{0.4cm} $[cm]$&$10\pm 0.001$ & $2\pm 0.01$ \\
$M\pm \Delta M$ \hspace{0.2cm}$[kg]$& $10 \pm 10^{-5}$ & $10\pm  10^{-7}$\\
\hline
$\boldsymbol{S}\pm \boldsymbol{\Delta S}$&  $<0\pm \boldsymbol{10^{-4}}$  & $<0\pm \boldsymbol{10^{-9}}$\\
\hline
\end{tabular}
\caption{Proposed experimental parameters to constrain the EAP violation parameter $S$ through the cavendish setup  as outlined in the  figure \ref{Cavendish} a). Case i values are similiar to what has been used in regular Cavendish tests \cite{EOTWAHPhysRevLett.85.2869}, while case ii shows values for picorad/s$^{2}$ sensitivity. 
The bounds can beat current laboratory bounds by many orders of magnitude \cite{Kreuzer}, enabling the exploration of the EAP for varying materials.}
\label{table:modified_cavendish}
\end{table}

We can also rewrite eqs. \eqref{eq:forces} in terms of the center-of-mass coordinate $\boldsymbol{X}_{cm}=(m_{1p}\boldsymbol{x}_1+m_{2p}\boldsymbol{x}_2)/M$, where $M=m_{1p}+m_{2p}$, resulting in its equation of motion 
\begin{equation} \label{eq:forcesCOM}
M \ddot{\boldsymbol{X}}_{cm} = \boldsymbol{F}_{net}
\end{equation}
with the net force
\begin{equation} \label{eq:Netforce}
\boldsymbol{F}_{net}=S \boldsymbol{F}_N
\end{equation}

This shows a breakdown of Newton's third law as momentum of the joint system is no longer conserved. However, despite the seemingly radical breakdown of normal physics, this net force can be challenging to detect. The dynamics still follows Newton's regular 2nd law: $\boldsymbol{F}=m_i \boldsymbol{a}$. As long as the normal equivalence principle holds, $m_p=m_i$, the passive mass thus completely drops out from the equations of motion generated by \eqref{eq:forces} or equivalently \eqref{eq:Netforce}. Thus, despite the EAP violation, gravitating systems follow the regular rules of Newtonian gravity, only with the relevant masses replaced by $m_a$. Without any direct knowledge of $m_p$, there are thus no modifications to Kepler's laws or other gravitational effects. On the level of the joint degrees-of-freedom, the crux is that the center-of-mass is no longer unique. Rather than using the passive mass as in eq. \eqref{eq:forcesCOM}, one can define the active center-of-mass $\boldsymbol{X}_{cm}^a$ with respect to only $m_a$, for which eq. \eqref{eq:forcesCOM} becomes $M_a \ddot{\boldsymbol{X}}_{cm}^a = 0$.  This means, and one can show this explicitly (see Appendix A), that the composite system does not self-accelerate with respect to a third body: the mutual relative distances remain unaltered by the EAP violation. A purely gravitational experiment thus cannot determine the EAP \cite{TrederGalillei} -- unless $m_p$ is measured independently as in the case of our proposed Cavendish-type setup in Fig.\ref{Cavendish}.

There is, however, a way to probe the anomalous net-force \eqref{eq:Netforce}. To this end, one requires additional interactions other than gravity. In this way $m_p$ does not cancel from the dynamic equations and thus the $S$-parameter can be probed through detection of the net-force. Before describing such a scenario, let us first highlight one additional anomaly in the purely gravitational case. The equations \eqref{eq:forces} have two free parameters: $m_{1a}/m_{1p}$ and $m_{2a}/m_{2p}$. Thus the single $S$-parametrization used so far in the literature, eq. \eqref{eq:S}, is insufficient to capture all violations of the EAP even in the classical case. We can see that explicitly: in addition to the CM-degree of freedom \eqref{eq:forcesCOM} we also have for the relative degree of freedom $\boldsymbol{x}\equiv \boldsymbol{x}_1-\boldsymbol{x}_2$ and reduced mass $\mu^{-1} =  m_{1p}^{-1} + m_{2p}^{-1}$:
\begin{equation} \label{eq:Frel}
 \mu \ddot{\boldsymbol{x}} = (\sigma - 1) \boldsymbol{F}_{N}
 \end{equation}
with a new parameter that arises for this relative motion:
\begin{equation} \label{eq:sigma}
    \sigma =  \frac{m_{1p}+ m_{2p} - m_{1a} - m_{2a}}{m_{1p} +m_{2p}} = 1 - \frac{m_{1a} + m_{2a}}{m_{1p} + m_{2p}} \, .
\end{equation}
This parameter has so far been overlooked in the literature and there are no dedicated constraints on it. But from eq. \eqref{eq:Frel} it is clear that its effect is to change the strength of the normal, mutual gravitational force between the two particles. There can be instances where $S=0$ but still an EAP violation with $\sigma \neq 0$. From current precision measurements of $G$ \cite{LutherTowlerPhysRevLett.48.121,QuinneetalPhysRevLett.87.111101,EOTWAHPhysRevLett.85.2869,BigG:ArmstrongPhysRevLett.91.201101,QuinnParksSpeakeBIG:G, westphal2021measurement} we get the bound $\sigma \lesssim 10^{-4}$. In fact, the discrepancies of measurements of $G$ across different experiments would be consistent with a weak EAP violation. The existence of this $\sigma$-parameter thus motivates dedicated experiments that compare the values of $G$ for varying materials -- in particular for varying the massive source masses, as they are taken to be just tungsten, lead and stainless steel in most experiments to date \cite{rothleitner_schlamminger_g}.  Fig. \ref{Cavendish} b) shows the Cavendish setup that is only sensitive to $\sigma(M1,M2)$, and not $S$.

We now focus on the detection of $\boldsymbol{F}_{net}$ as an alternative test of EAP violations. This method was first realized and utilized by Bartlett and van Buren (BvB) in 1986 \cite{Bartlett}, putting stringent bounds on $S$ from lunar motion. Here we will show that it not only extends into laboratory settings, but that it also enables tests of unique quantum violations on the atomic scale. As mentioned above, in purely gravitational interactions the relative motion of systems is qualitatively the same as in regular physics. But once other forces come into play, the breakdown of Newton's third law becomes relevant. We generalize \eqref{eq:forces} to include a general non-gravitational force $\boldsymbol{F}^{non-gr}$, for which the usual laws of physics hold\footnote{For considerations and experimental constraints on violations of the equality of active and passive electric charges, see Ref. \cite{Muller.A-P.electric.charge}.}. This yields the CM dynamics
\begin{equation} \label{eq:forcesCOMnonG}
    \ddot{\boldsymbol{X}}_{cm} =S\boldsymbol{F}_{21}^{non-gr} \, .
\end{equation}

We note here that even the notion of `center-of-mass' becomes ambiguous. One can define it with respect to $m_p$, $m_a$ or any combination of them. 
This issue, however, is fully avoided in the extreme case of a rigid composite system. In this case, $\boldsymbol{F}^{non-gr}$ becomes a constraint force. The limit of a perfectly rigid composite system is useful also because the nature and strength of the non-gravitational force becomes irrelevant for our considerations of the EAP. Instead, the entire system now accelerates according to
\begin{equation} \label{eq:COMnonG}
    \ddot{\boldsymbol{X}}_{cm} = S G \mu\frac{\boldsymbol{x}}{|\boldsymbol{x}|^3} 
\end{equation}
\begin{figure}[b]
 \centering
 \begin{adjustbox}{center}
   \includegraphics[width=0.8\columnwidth]{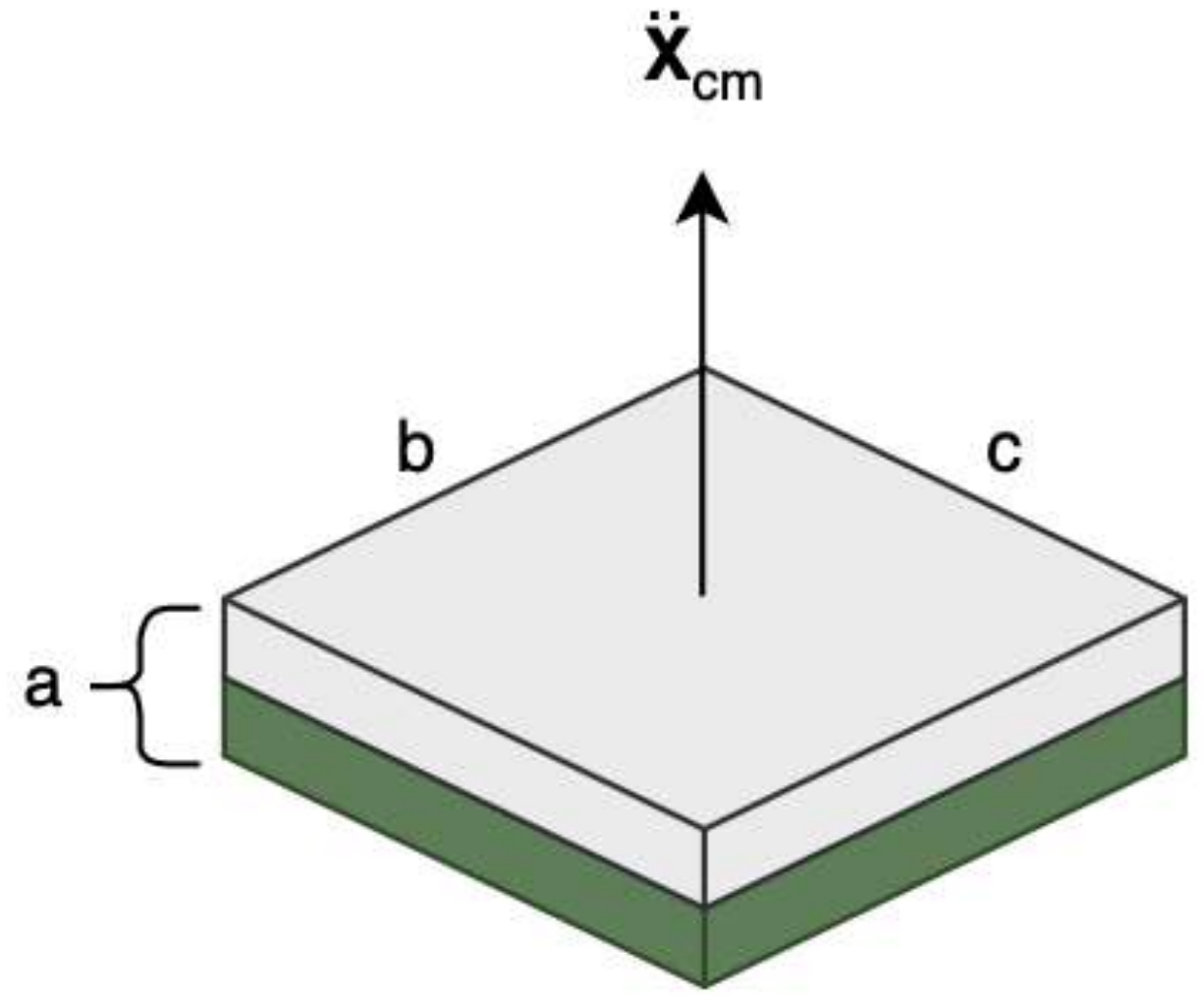}
 \end{adjustbox}
 \caption{ Two uniformly distributed thin films attached together, each made of a different material. The total volume is $V\equiv abc$. For simplicity, we assume that each of the thin films, $1$ and $2$, have identical dimensions. Therefore, \(V_1 = V_2 = \frac{1}{2}abc\). If there is a non-zero $S$-parameter between the two materials, then the system should self-accelerate, allowing for a precise test of the EAP.
 } 
 \label{Figure:Slab}
\end{figure}

\begin{table}[t]
\centering
\begin{tabular}{ |p{2.8cm}||p{2.2cm}|p{2.2cm}|  }
 \hline
 \multicolumn{3}{|c|}{Thin slab free falling in space} \\
 \hline
\hspace{0.6cm}Variables & \hspace{0.5cm} Case 1 & \hspace{0.5cm} Case 2\\
 \hline
\hspace{0.2cm} $|\ddot{\boldsymbol{X}}_{cm}|$ \hspace{0.1cm} $[fm/s^2]$ &  \hspace{0.8cm} $1$    & \hspace{0.8cm} $1$ \\
\hspace{0.5cm} $\rho_1$ \hspace{0.4cm} $[kg/m^3]$&   \hspace{0.5cm}$1.9\cdot 10^4$  & \hspace{0.4cm} $1.9\cdot 10^4$   \\
 \hspace{0.6cm}$\rho_2$\hspace{0.5cm} $[kg/m^3]$& \hspace{0.4cm} $2.1\cdot 10^4$ & \hspace{0.4cm} $2.1\cdot 10^4$ \\
\hspace{0.5cm} $a$ \hspace{0.65cm}$[\mu m]$& \hspace{0.8cm}$10$ & \hspace{0.7cm} $ 10^{-3}$ \\
\hspace{0.6cm}$b$ \hspace{0.7cm}$[m]$& \hspace{0.9cm}$1 $ & \hspace{0.9cm}$1$\\
\hspace{0.6cm}$c$ \hspace{0.7cm}$[m]$&\hspace{0.8cm} $1$ &\hspace{0.8cm} $1$\\
\hline
\hspace{1.3cm}$\boldsymbol{S} $ & \hspace{0.4cm}  $<\boldsymbol{10^{-14}}$  & \hspace{0.4cm} $< \boldsymbol{10^{-18}}$\\
\hline
\end{tabular}
\caption{ The first column lists all relevant physical quantities, including the dimensions and densities of the films. For all cases, we consider the films are made of some high density material. We choose gold and platinum,
as in \cite{LISAPhysRevLett.116.231101} and the acceleration resolution achieved by LISA pathfinder \cite{LISAPhysRevLett.116.231101}. Case $1$ competes with the constraints set by \cite{Singh}. In particular, given the acceleration precision of \cite{LISAPhysRevLett.116.231101}, we find the optimal material and size that could reach current bounds on S set by Ref. \cite{Singh}. Case $2$ can potentially outperform LLR experiments, given that such thin bi-layer systems can be fabricated.}
\label{table:slab}
\end{table}
The introduction of internal non-gravitational forces thus reveals the breakdown of momentum conservation, and an anomalous self-acceleration can be observed. BvB applied this principle to the moon, which they modeled as consisting of an Al-mantle and Fe-core based on Refs. \cite{Moon.model.asymmetry.1972apollo,muller1972moon,2-component-Moon.Wood1973}. Assuming these two parts of the moon are held by non-gravitational forces, an acceleration relative to Earth should therefore take place. Lunar laser ranging was used to constrain this self-acceleration and thus the parameter $S$ \cite{Bartlett}, most recently to $S \leq 3.9\cdot 10^{-14}$ \cite{Singh}. This remains the best bound on EAP violations to date. 

However, the principle of self-acceleration can be extended beyond celestial dynamics. Intuitively, one would expect that only large masses allow one to observe gravitational effects that involve sources of gravity. But as we saw above, an EAP violation results in a self-acceleration force which can become substantial even on microscopic scales. We can exploit these insights to design laboratory experiments which can probe EAP in unprecedented domains. The key ingredient we outlined above is that the system must be held together by non-gravitational forces. For example, one can consider the self-acceleration of composite systems such as nuclei (self-force between neutrons and protons), molecules, rotating Janus-particles made of different materials \cite{JanusreviewPS,GHz.Rotation.trapped.nanoparticles}, or levitated sphere \cite{westphal2021measurement}, to name a few. For a Deuterium nucleus, for example, the magnitude of the self-acceleration would be $|\ddot{\boldsymbol{X}}_{cm}|=SG\mu \frac{1}{r^2}= 6\cdot 10^{-8}S~ [\frac{m}{s^2}]$, where $\mu$ the reduced Deuterium mass and $r$ the distance between proton and neutron. However, such systems typically rotate and thus the displacement would average out.
On average, the anomalous acceleration would therefore not be observable (we note that this in fact also applies to the case of the moon, however, since it is tidally locked to the Earth there is no rotation with respect to the Earth and thus the BvB experiment can constrain $S$.) In addition, the self-acceleration increases as the number of constituents increases. Thus we propose the setup described in Figure \ref{Figure:Slab}. Two attached thin films made of different materials, each with a uniform mass distribution. These films can be fabricated to be extremely thin approximating two-dimensional sheets \cite{GOLDTHICKKossoy_2014}. The self-acceleration in this case becomes:
\begin{equation}
   |\ddot{\boldsymbol{X}}_{cm}| = SG \frac{\rho_1 \rho_2}{\rho_1 + \rho_2} \frac{bc}{a}
\end{equation}
Here, $\rho_1,\rho_2$ represent the densities of the materials, $b,c$ denote the length and width respectively while $a$ corresponds to the thickness, as shown in Figure \ref{Figure:Slab}. Table \ref{table:slab} summarizes some numerical estimates. They show that through precise measurements of acceleration, such as in space-based missions \cite{LISAPhysRevLett.116.231101}, the self-acceleration of this system could be detected or the EAP parameters constrained to very high precision.

\section{Quantum violations of the EAP}
We now turn to the EAP in the quantum domain which can serve as a new probe of quantum gravity phenomenology, both directly and indirectly. In some models, such as for quantum foam, a breakdown of Newton's third law is postulated \cite{Bekenstein_2014}. Even in string theory the mass parameter appearing in the different resulting metric components $g_{00}(M)$ and $g_{ij}(M')$ might vary \cite{garfinkle1991charged}, which also amounts to violations of the EAP \cite{vollick2022meaning}. But even without specific models, testing how quantum systems source gravity is of central interest, as the resulting gravitational fields would have to be described quantum mechanically. There are now several proposals to create spatial superpositions of gravitating masses \cite{GIEBose,GIEMarlettoVedral,carlesso2019testing,krisnanda2020observable,anastopoulos2020quantum,pedernales2022enhancing,kaku2023enhancement,hanif2024testing}, but these remain very challenging to implement. We now show that EAP tests in the quantum domain can offer an alternative route. Unique to gravity is the fact that even the gravitational charge itself must be quantized. The stress-energy tensor that captures quantum matter and energy sources the field via $G_{\mu \nu} = 8 \pi G T_{\mu \nu}/c^4$. There is currently no empirical evidence how a quantum mechanical source mass would produce the gravitational field, and there are many speculative alternatives such as semi-classical gravity sourced only by $\langle \hat{T}_{\mu \nu} \rangle$ \cite{singh1989notes}. In other words, probing whether gravity can be sourced by operator-valued charges tests physics deep in the quantum gravity regime.  Here we will focus on the limit of a composite low-energy system that is well captured by the rest-mass $m$ but that also can have addressable internal energy excitations $E_i$, describing for example an atom with electronic or nuclear energy levels or a trapped nanosphere with embedded addressable spins.  
Such a description is well-suited for low-energy quantum systems in weak-field general relativity \cite{pikovski2017time, PikovskiGravitationalMass, schwartz2019post, martinez2022ab}. Since this energy is quantized, thus also the total relevant mass must be a quantum operator $\hat{M}=m+\hat{E}/c^2$ \cite{zych2018.QUANTUM.EP, zych2011quantum, pikovski2017time}. For the passive mass, this is an experimental fact at low energies: all atomic clock experiments follow precisely this principle when they record relativistic time dilation through superpositions of atomic energy levels \cite{Hafele1972, Chou2010,Zheng_2022,Ludlow2015}, as the energy couples to the external gravitational field and produces the redshift. 
In contrast, the quantum nature of the active mass is far from given, and such a quantum source of the gravitational field would require a full quantum gravity description. Needless to say, probing the quantum nature of a gravitational source mass directly would be extremely challenging. 
A test of the EAP for mass superpositions is an alternative route -- which can provide an indirect test of the same quantum gravity principles. Namely, if quantum sources of gravity do not follow the usual quantum rules that apply to probe particles, then an EAP violation must occur. Tests of the EAP in the quantum domain thus open new indirect hints of new physics related to the interplay of quantum theory and gravity deep in the quantum gravity domain. We now show that one can probe and constrain such quantum EAP violations with AMO systems, and thus test to what extent quantum source masses (which are very hard to create) could behave differently from quantum probe masses (which are routinely used experimentally).

Taking mass-energy equivalence and quantum superpositions of energy into account, the EAP at the operator level becomes
\begin{equation}\label{A-P.equalityOperators}
\hat{M}_a=\hat{M}_p    
\end{equation}
where $\hat{M}_j=m_j\hat{I}_{int}+c^{-2}\hat{E}_{int,j}$ for $j=a,p$ and the subscript highlights that these are internal energy levels, which we omit in the following. 
A classical EAP, $m_a=m_p$, is insufficient to  imply the equality of the corresponding mass operators. In other words, it is conceivable that even though EAP could be exactly valid in the classical domain, in the quantum domain it may be violated for example if superpositions are involved. This is in analogy to the equivalence principle, where a purely quantum mechanical violation is possible \cite{rosi2017quantum,zych2018.QUANTUM.EP}. To test such quantum EAP violations, we use the lessons from the previous section focusing on classical dynamics. While a Kreuzer-type experiment would require the exact measurement of passive mass even for individual energy levels, and is thus not feasible, the other type of test outlined above is well suited for the quantum domain: testing for a self-acceleration. For two localized gravitating systems, their mutual forces that may be EAP violating are now expressed as follows
\begin{equation} \label{eq:quantumforces}
\begin{split}
    \hat{F}_{12}  &=-G  \hat{M}_{1a}\hat{M}_{2p}\frac{1}{r^2}   \\
  \hat{F}_{21}  & =G  \hat{M}_{2a}\hat{M}_{1p}\frac{1}{r^2}  
\end{split}
\end{equation}
with the net force given by\footnote{In what follows, the quantum formulation of the weak equivalence principle is assumed to be valid, $\hat{M_i}=\hat{M_p}$}.
\begin{equation}\label{Fnetquantum}
  \hat{F}_{net}\equiv(\hat{M}_{1p}+\hat{M}_{2p})\hat{\ddot{X}}_{cm}=-\frac{G}{r^2} (\hat{M}_{1a}\hat{M}_{2p}-\hat{M}_{1p}\hat{M}_{2a}) .
\end{equation}
Here we assume the relative distance $r$ to be fixed between the systems, but that the joint center-of-mass can accelerate as described by $\hat{\ddot{X}}_{cm}$, which in general will be an operator since the force depends on the operator-natured masses. We highlight however, that here we can only work on the level of equations of motion (there is no Hamiltonian formalism that can capture EAP violations), and we are interested in finding the resulting acceleration of the particle, again assuming a composite system constrained by other forces.

By explicitly expanding the active and passive mass operators using the quantum formulation of mass-energy equivalence, the acceleration operator can be expressed as a series comprising various terms, which can be grouped according to powers of $1/c$. This approach, schematically, yields 
\begin{equation}
\hat{\ddot{X}}_{cm}\equiv \sum_{n=0}^2\frac{1}{c^{n}}\hat{\ddot{X}}^{(c^{-n})}_{cm}  
\end{equation}
The leading-order term in the expansion, $\ddot{X}^{(c^0)}_{cm}$, corresponds to the self-acceleration, as given by \eqref{eq:COMnonG}, while the next to leading order term reads as\footnote{To derive \eqref{self-acc(c minus two)}, we assume that the system's rest mass is much larger than the addressable internal energy.}
\begin{equation}\label{self-acc(c minus two)}
\begin{split}
\hat{\ddot{X}}^{(c^{-2})}_{cm} & = -\frac{G}{r^2}\mu_p\biggl[\frac{m_{1a}+m_{2a}}{m_{1p}+m_{2p}}\biggl(\frac{\hat{E}_{2p}}{m_{2p}c^2}-\frac{\hat{E}_{1p}}{m_{1p}c^2}\biggr) \\ 
& \quad +\frac{\hat{E}_{1a}}{m_{1p}c^2}  -\frac{\hat{E}_{2a}}{m_{2p}c^2}\biggr] \, .
\end{split}
\end{equation}
We consider the scenario $S=0$ in which no violations occur for the rest masses, meaning the classical EAP $m_a=m_p$ holds, while quantum violations could arise due to superpositions of internal energy states. Specifically, we focus on situations where one system, referred to as the ``clock'' system which has internal energy superpositions, is attached to another system without quantized internal energy levels. In this case, eq. \eqref{self-acc(c minus two)} simplifies to give rise to an expression for the self-acceleration of the entire system
 \begin{equation} \label{eq:quantumCM}
\hat{\ddot{X}}^{(c^{-2})}_{cm}= -\frac{G}{c^2 r^2}\frac{m_{2p}}{m_{1p}+m_{2p}} \left( \hat{E}_{1a} -\hat{E}_{1p} \right)
\end{equation}   
This is the central equation that allows us to explore quantum EAP violations. Importantly, this equation shows that we can now directly probe the nature of the quantized active source energy: the self-acceleration is effectively measuring $\hat{E}_{1a}$ against $\hat{E}_{1p}$. But we can know the passive energy of the system very precisely through clock measurements. Thus it provides a ``ruler'' to measure the otherwise inaccessible active energy, and any discrepancy between the two would result in a self-acceleration. It is an operator because we use the quantum formalism, but in general it could be replaced by some other quantity or function, depending on the model for how quantum energy sources gravity. For example, violations could occur when superpositions arise and it might be the case that $\hat{E}_{1a}=0$ when superpositions are involved, or that $\hat{E}_{1a} \rightarrow \langle \hat{E}_{1a} \rangle$. In both cases, the self-acceleration would consist of two different accelerations in superposition, while the regular EAP case would of course result in no self-acceleration at all. Another possibility is that $\left[ \hat{E}_{1p} , \hat{E}_{1a} \right] \neq 0$, which could result in oscillations of active or passive mass in analogy to neutrino oscillations and thus cause an oscillatory or fluctuating self-acceleration. Without knowing the detailed nature of the EAP violation and the model, we can still simply parametrize any such occurrence by introducing a new parameter $\hat{S}_q$: 
 \begin{equation} \label{eq:quantumCMSq}
\hat{\ddot{X}}^{(c^{-2})}_{cm}= -\frac{G}{ r^2}\frac{m_{2p}}{m_{1p}+m_{2p}} \frac{ \langle \hat{E}_{p}\rangle }{c^2}\hat{S}_q
\end{equation}   
We choose to normalize the violations by $\langle \hat{E}_{p}\rangle$ since these are the degrees of freedom and energy scales on which we are probing the EAP. For example, for a two-level system this would be simply the transition energy $\hbar \omega$. The new parameter $\hat{S}_q$ can be quite general since we can leave the nature of the active energy open, and it may not even be an operator as in normal quantum theory as it may also depend on the internal states (such as for example when superpositions do not source gravity according to quantum rules). But in terms of the regular quantum description we would have $\hat{S}_q= (\hat{E}_{a} -\hat{E}_{p})/\langle \hat{E}_{p}\rangle$. We emphasize that a test of $\hat{S}_q$ dives deep into possible new physics in the quantum gravity realm. A property of a quantum source mass that would generate gravity in a fashion inconsistent with how a quantum probe particle feels gravity would show up as a nonzero $\hat{S}_q$ parameter. Thus its test is also a test whether gravity can be sourced by quantized matter.

\begin{figure}[b]
 \centering
 \begin{adjustbox}{center}
   \includegraphics[width=0.9\columnwidth]{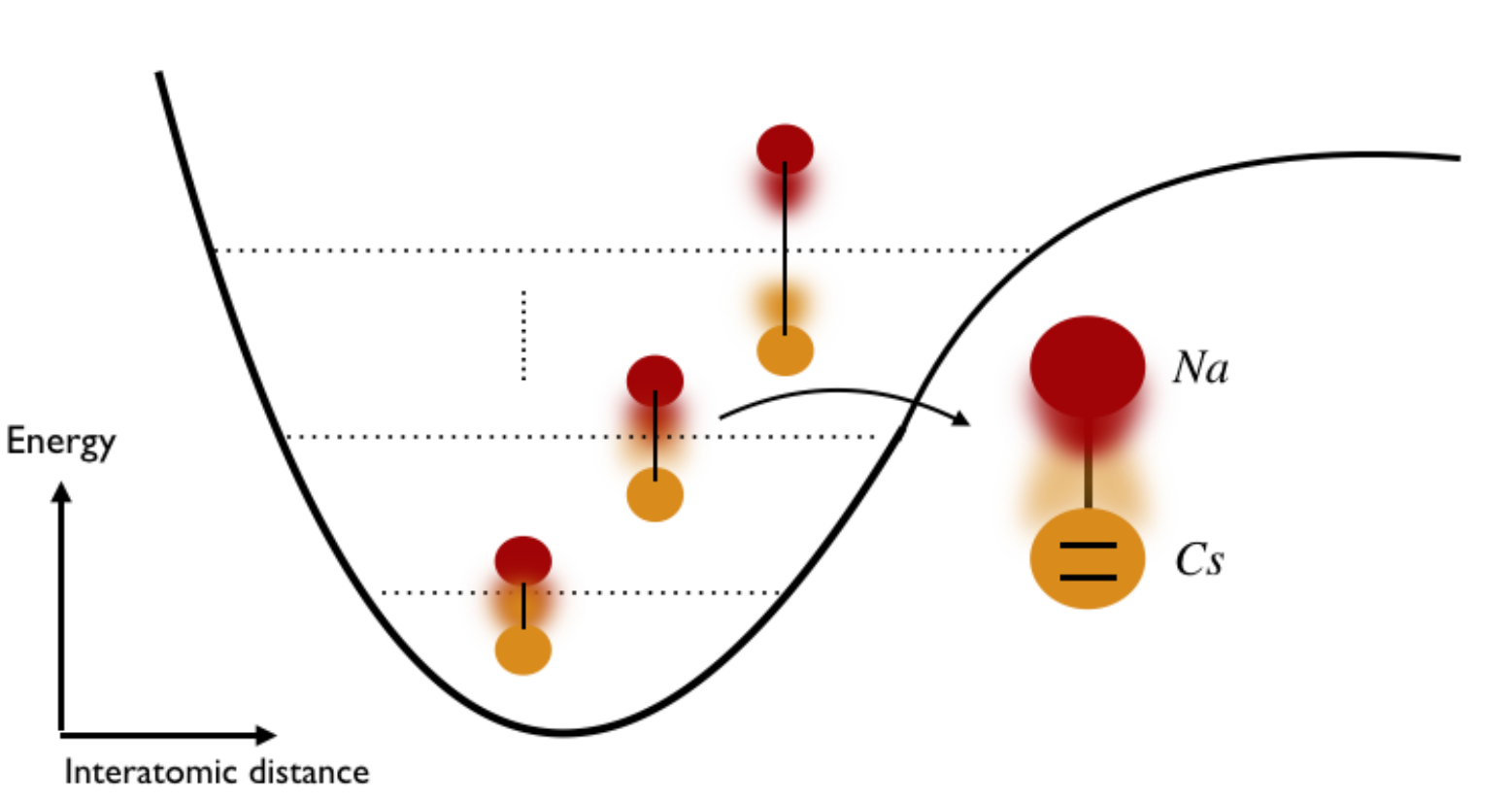}
 \end{adjustbox}
 \caption{Schematic of weakly, intermediate, and tightly bound Na-Cs molecules. In the tightly bound case, the atoms are in close proximity, with a significant overlap between their electronic wavefunctions. For weakly bound molecules, the interatomic distance is large, resulting in minimal overlap. Of particular interest is the intermediate case, where the atoms are close but where the orbital of each is still well-localized. Preparing a superposition in the Cs-atom thus allows one to probe EAP violations due to quantum superpositions, by probing any resulting self-acceleration as in eq. \eqref{eq:quantumCMSq}.}
 \label{NaCs}
\end{figure}

\begin{figure}[t]
 \centering
 \begin{adjustbox}{center}
   \includegraphics[width=.95\columnwidth]{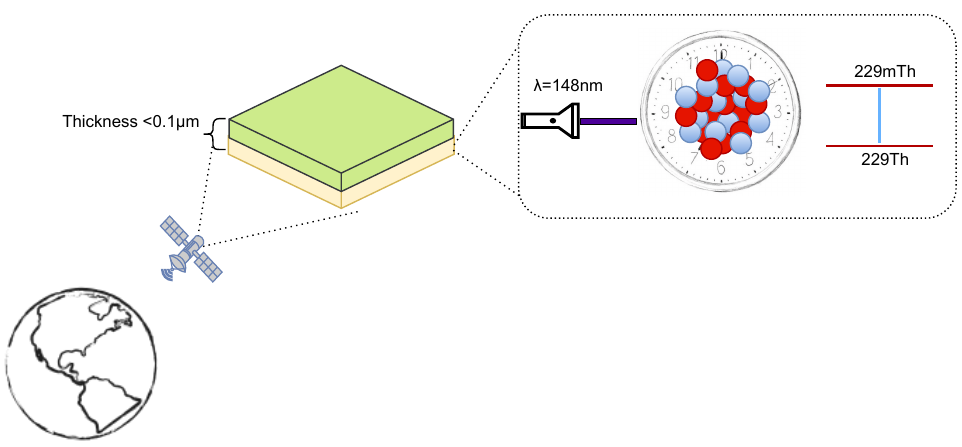}
 \end{adjustbox}
 \caption{Schematic of the proposed experiment to test quantum violations of the EAP with nuclear clocks. The lower layer of the material, shown in yellow, is doped with $N$ Thorium nuclei prepared in energy superposition states, while the upper layer consists of a material without any excited Thorium nuclei. Given the acceleration resolution that can be achieved in space \cite{LISAPhysRevLett.116.231101}, constraints $S_q \lesssim 1$ could be achieved with $N \sim 10^{16}$ clocks.}
 \label{NuclearClock}
\end{figure}

We now consider the experimental scenarios that can constrain $\hat{S}_q$. First, we emphasize the requirements for such tests based on the obtained results: We need two systems held together by non-gravitational forces, one of the masses to have addressable energy levels to create superpositions of energy, the other mass having $m_{2p} \gtrsim m_{1p}$ due to the rest mass ratio in \eqref{eq:quantumCMSq}, and the distance between the two masses as small as possible. Secondly, we highlight that the exact EAP would yield no acceleration at all. The experiments to constrain the self-acceleration can thus give upper bounds on any of the possible eigenvalues $\lambda$ of $\hat{S}_q$, which we denote by $S_q = \max_{ \{ \lambda \} } \hat{S}_q$. We don't need to probe the quantum nature of the acceleration per se, but rather probe whether any acceleration can arises if quantum states are involved.
This leads us to two possible tests with AMO technology as examples: with cold polar molecules (figure \ref{NaCs}), and with nuclear clocks (figure \ref{NuclearClock}). We use these two examples to illustrate the ability to constrain $S_q$.

We first consider an AMO implementation with cold polar molecules. We require that the superposition in one mass is well-localized such that the net-acceleration is not averaged out. This means the system should also not be rotating, and the energy levels need to be associated with well-defined spatial regions.
Natural candidates for superpositions in the quantum domain, such as two different nuclear levels,  nuclear and electronic levels or even several electronic transitions \cite{hohn2024determining}, are therefore not ideal if they average out the net-force. Instead, we suggest the setup depicted in Figure \ref{NaCs}, based on recent advancements in ultra-cold polar molecules \cite{chomaz2022dipolar}. Polar molecules seem to be good candidates as they consist of two different atomic species, their interatomic distance can be tuned through Feshbach resonances \cite{inouye1998observation}, they can be cooled to the ground state \cite{bigagli2024observation} and they can be aligned through an external magnetic field. We consider the self-acceleration of a dipolar $\text{Na-Cs}$ molecule as recently demonstrated as a BEC \cite{bigagli2024observation}, with the $\text{Cs}$ atom prepared in a superposition of different energy eigenstates. We require that the two atoms are sufficiently far apart such that the electronic levels do not fully overlap and the two superpositions are spatially distinct. This requires the binding, and thus the distance between the molecules to be relatively loose. At the same time, we do not want the molecule to disassociate under laser interrogation of the levels (here we require only superposition state preparation and not a full Ramsey read-out sequence). For this, we compute the overlap function $\mathcal{S}_{\text{Na-Cs}}=\int dV \Psi^{*}_{\text{Na}}\Psi_{\text{Cs}}$ for the \text{Na-Cs} molecules (see Appendix D). Requiring the molecule to have binding energy of order $\text{GHz}$ and overlap $\mathcal{S}_{\text{Na-Cs}}\lesssim 15\%$,  amounts to an optimal separation of the two atoms of $r\approx 10 \r{A}$. 
 We now estimate the constraints that can be placed on $S_q$ with this system. The rest masses are $m^{\text{Cs}}_{p}\approx 2 \times 10^{-25} ~\text{kg}$, $m^{\text{Na}}_{p}\approx 4 \times 10^{-26} ~\text{kg}$. We also consider superpositions of $\Delta E_{p}^{\text{Cs}}/c^2\approx 5\cdot 10^{-36} ~\text{kg}$, the equivalent mass of an optical energy superposition at $\lambda =414$~\text{nm}. 
If we assume a ground-based experiment where ultracold NaCs polar molecules are released into free fall,  assuming an acceleration measurement precision of $a \sim 10^{-10}~$m/s$^2$ (similarly to what has been achieved with atomic fountains \cite{asenbaum2020atom}), we get the ability to constrain ${S}_{q}$ from \eqref{eq:quantumCMSq} if no anomalous self-acceleration is observed along the axis of the dipolar system. With these numbers we would get a constraint of ${S}_{q} \lesssim 6 \times 10^{17}$.  
A possible differential experiment could then be using a trapped Na-Cs-BEC as demonstrated in ref. \cite{bigagli2024observation} with half of the polar molecules prepared in internal superposition states of the Cs, and then released from the trap to compare their free-fall displacement to those without any internal superposition. An EAP violation would thus show as a seeming violation of the usual equivalence principle, the molecules with and without the superpositions would fall at different rates. Similarly, one can envision such an experiment in space to significantly increase the free-fall times, to say $t=100$~s and probe $S_q$ to a greater precision. A two-species BEC has in fact been proposed for space missions \cite{BEC2speciesinspace}, using $K^{41}$ and $Rb^{87}$. If a molecular BEC is prepared with the atoms bound together, a stringent test of possible quantum violations of the EAP could then be performed in space. Using as an example the acceleration imprecision of LISA pathfinder \cite{LISAPhysRevLett.116.231101} one would constrain  $S_q \lesssim 10^{12}$.

Another possibility is a test with nuclear clocks, such as Thorium clocks whose 8eV transition has recently been demonstrated \cite{tiedau2024laser,elwell2024laser,zhang2024frequency}. The great advantage of using nuclear clocks is the ability to use $N$ clocks in one sample, and we can combine this capability with the bi-layer material self-acceleration probe as described in the section on classical tests. Importantly, this scales up the self-acceleration in eq. \eqref{eq:quantumCMSq} by a factor $N$ since now $N$ systems in superposition are involved. Concretely, assume we have a solid-state host for the Thorium, like a CaF$_{2}$ calcium fluoride crystal. We prepare the Thorium in only half of the material in superposition states, localized in one area, as in figure \ref{NuclearClock}. The interaction with the other half of the material will then result in a self-acceleration if an EAP is present, according to eq. \eqref{eq:quantumCMSq}. The rest-mass of the Thorium clocks will be less than that of the host crystal on other end, and we approximate the distance between the clocks and the other part of the material as the distance of the center-of-masses of the two sides of the crystal. We assume we can prepare the energy transition of $148~$nm for $^{229}$Th in superposition for $N=10^{16}$, which is the goal of nuclear clocks. For a $^{229}$Th density of $\rho \sim 10^{22}$~atoms/cm$^{3}$ we need a clock material volume of $V\sim 10^{-12}$~m$^{3}$. 
If we assume a mm-sized thin film of $~0.1$~$\mu$m thickness giving $r \sim 10^{-7}$~m for the relevant interaction distance with the other part of the film without the Thorium clocks. Again assuming the LISA pathfinder acceleration imprecision $a < 1~$fm/s$^2$, such a system can then constrain (or detect) the self-acceleration to $S_q \lesssim 1$. This shows that future nuclear clock technology, combined with tests of Newton's third law, enable new fundamental tests such as the quantum EAP, and remarkably, they can probe violations deep in the quantum gravity regime where differences between active and passive quantized energies would show up even if they were of order unity.

We stress that there are no a priori values of $S_q$ that are constrained and thus such violations may occur for values much larger that unity. Dedicated experiments as described above, or other possible implementations which involve energy superpositions, would provide unprecedented bounds, and a new test of the interplay between quantum and general relativistic foundational principles. But pushing into the domain where $S_q \lesssim 1$ is of most interest, as this allows one to directly test how quantized energy sources gravity through $\hat{E}_a$, by using the well-known passive energy $\hat{E}_p$ as a point of reference.

\section{Conclusions}
In this work we have focused on an under-explored fundamental principle of gravity: the equivalence of passive and active gravitational mass. While being a critical pillar of our current understanding of physics, it  relates to the subtle interplay of purely relativistic effects where energy contributes to mass, and purely quantum mechanical effects where energy is also quantized. Exploring this principle in the classical and quantum domains thus shines light on the foundations of general relativity and how gravity and quantum theory intertwine. Our results highlight how the EAP can be probed, and that alternative and untested violations can be tested in laboratory experiments. Pure quantum violations offer a new route to indirectly probe quantized energy as a source of the gravitational field, and thus quantum gravity. In addition, some models of quantum gravity directly predict EAP violations. 
We outline how such violations can be detected by exploiting the necessary breakdown of Newton's third law, which enable tests with microscopic, atomic and novel clock systems in ground- and space-based experiments. Our results show new opportunities for foundational tests of general relativity, its interface with quantum mechanics and possible signatures of new physics. 

\section{acknowledgments}
We thank Konstantin Beyer for discussions on quantum violations of the EAP, Michael Kopp for discussions on EAP violations and Sebastian Will for discussions on experiments with polar molecules. This work was supported by the Swedish Research Council under grant no 2019–05615, the NSF under grant No 2239498, NASA under grant No 80NSSC25K7051 and the Sloan Foundation under grant No G-2023-21102. 


%

\appendix
\section{Appendix A: Gravitational dynamics with respect to the Active Center of Mass}
For three mutually gravitating masses, starting from Newton's equations of motion, after setting $m_i=m_p$, we get
\begin{align} \label{threebody1}
    \ddot{\boldsymbol{x}}_1& = -Gm_{2a}\boldsymbol{\nabla}_1\frac{1}{|\boldsymbol{x}_1-\boldsymbol{x}_2|}  -Gm_{3a}\boldsymbol{\nabla}_1\frac{1}{|\boldsymbol{x}_1-\boldsymbol{x}_3|} \\ \label{threebody2}
     \ddot{\boldsymbol{x}}_2& = -Gm_{1a}\boldsymbol{\nabla}_2\frac{1}{|\boldsymbol{x}_1-\boldsymbol{x}_2|}  -Gm_{3a}\boldsymbol{\nabla}_2\frac{1}{|\boldsymbol{x}_2-\boldsymbol{x}_3|}  \\ \label{3bodyeom3}
     \ddot{\boldsymbol{x}}_3& =-Gm_{1a}\boldsymbol{\nabla}_3\frac{1}{|\boldsymbol{x}_1-\boldsymbol{x}_3|}  -Gm_{2a}\boldsymbol{\nabla}_3\frac{1}{|\boldsymbol{x}_2-\boldsymbol{x}_3|}.
\end{align}
In order to describe dynamics of the total system, we define the relative distance $\boldsymbol{x}_{12}=\boldsymbol{x}_2-\boldsymbol{x}_1$ between objects $1$ and $2$ and their active center of mass as $\boldsymbol{X}^{a}_{cm}=(m_1^a\boldsymbol{x}_1+m_2^a\boldsymbol{x}_2)/(m_1^a+m_2^a)$. The relative distance between the active CoM of the composite system $1,2$ and mass $3$ is $\boldsymbol{R}_{rel}=\boldsymbol{X}^{a}_{cm}-\boldsymbol{x}_3$. Using equations \eqref{threebody1},\eqref{threebody2} and setting the origin of the coordinate system at $\boldsymbol{x}_3=0$, we find that the active center of mass accelerates towards the third body 
\begin{equation}
\begin{split}
   |\ddot{\boldsymbol{X}}^{a}_{cm}| =-G\frac{m_{3a}}{m_{1a}+m_{2a}} \biggl[&\frac{m_{1a}}{|\boldsymbol{X}^{a}_{cm}-\frac{m_{2a}}{m_{1a}+m_{2a}}\boldsymbol{x}_{12}|^2}+\\
    & +\frac{m_{2a}}{|\boldsymbol{X}^a_{cm}+\frac{m_{2a}}{m_{1a}+m_{2a}}\boldsymbol{x}_{12}|^2}\biggr]
\end{split}
\end{equation}
This is equation is in agreement with standard results in the literature \cite{Gutzwiller3bodyRevModPhys.70.589}.
Assuming further that the distance of the third body is much larger than the separation of the composite system, namely $|\boldsymbol{X}^a_{cm}|\gg |\boldsymbol{x}_{12}|$, the above expression simplifies, to get a more familiar equation
\begin{equation}
|\ddot{\boldsymbol{X}}_{cm}^a|=-Gm_{3a}\frac{1}{|\boldsymbol{X}_{cm}^a|^2}    
\end{equation}
This equation corresponds to the standard equation of motion of a binary system towards a third mass within Newtonian gravity. Therefore, with respect to the active center of mass, celestial mechanics remains unchanged and EAP violations cannot be probed in a purely gravitational setting\cite{treder1983galilei}, unless $m_p$ is determined independently.

Nevertheless, celestial mechanics can be affected, if non-gravitational forces are also involved, as discussed in the main text. This is why the experiment by Bartlett and van Buren \cite{Bartlett} was able to constrain the S-parameter using lunar laser ranging. Assuming that the moon is a composite system that consists of two materials held together by non-gravitational forces, a net force will arise causing a net acceleration with respect to Earth.

\section{Appendix B: Cavendish experiments for constraining S}
\label{Appendix1}

Here, we illustrate a modified version of the Cavendish experiment aiming to put constraints on $S$. The setup we are considering is the leftmost in figure \ref{table:modified_cavendish} in the main text. This particular configuration is an adaptation that transforms the traditional Cavendish setup into a \textit{null experiment}. The primary objective of this modification is to isolate and analyze the signal (self-acceleration) that arises solely from a non-zero $S$ parameter between the two source masses $M_{1a}$ and $M_{2a}$. The combined influence of both source masses on the torsion balance is a total torque, as given by 
\begin{equation}
\boldsymbol{\tau}_{tot}\equiv \boldsymbol{\tau}_1+\boldsymbol{\tau}_2=-Gmd\frac{1}{R^2} (M_{1a}-M_{2a})\boldsymbol{e}_{\phi}  
\end{equation}
with
\begin{equation} 
\begin{split}
    \boldsymbol{\tau}_1  &=\boldsymbol{r}\times\boldsymbol{F}_{M_{1a}\rightarrow m}=-GM_{1a}md\frac{1}{R^2}  \boldsymbol{e}_{\phi}  \\
  \boldsymbol{\tau}_2  & =\boldsymbol{r}\times\boldsymbol{F}_{M_{2a}\rightarrow m}=-GM_{2a}md\frac{1}{R^2}(-\boldsymbol{e}_{\phi}) .
\end{split}
\end{equation} 
Expressing $M_{1a}$ and $M_{2a}$ in terms of $S_{12}$ and $\sigma_{12}$ as given by \eqref{eq:S} and \eqref{eq:sigma}, where the indices highlight that these are parameters referring to the two big source masses, the total torque takes the form
\begin{equation}\label{totaltorque}
\boldsymbol{\tau}_{tot}=-Gmd\frac{1}{R^2} \biggl[(1-\sigma_{12})(M_{1p}-M_{2p})-2\mu S_{12}\biggr]\boldsymbol{e}_{\phi}  .
\end{equation}
In this equation both $\sigma_{12}$ and $S_{12}$ appear. However, if the source masses have the same passive gravitational mass, $\Delta M_p\equiv M_{1p}-M_{2p}=0$  up to some experimental precision, which can be large and we quantify in table \ref{table:modified_cavendish}, the resulting total torque 
\begin{equation}
|\boldsymbol{\tau}_{tot}|=Gmd\frac{1}{R^2} M_p S_{12}  
\end{equation}
depends only on $S_{12}$. The corresponding angular acceleration of the torsion balance is 
\begin{equation}
    \ddot{\phi}=\frac{|\boldsymbol{\tau}_{tot}|}{I}=GM_p\frac{1}{dR^2}  S_{12} 
\end{equation}
with $I=md^2$ the moment of inertia for the torsion. Table \ref{table:modified_cavendish} provides some numerical estimates. \\\\

\section{Appendix C: Cavendish experiments for constraining $\sigma$}
\label{Appendix2}
Here, we now provide the mathematical details to describe the setup on the right in Figure \ref{table:modified_cavendish}. This setup is closer to a typical Cavendish experiment but with the source masses of different material. The resulting total torque exerted on the torsion balance, due to the presence of source masses $M_{1a}$ and $M_{2a}$, is 
\begin{equation}
    |\boldsymbol{\tau}_{tot}|=G\frac{md}{|\boldsymbol{r}_M-\boldsymbol{d}|^2} (M_{1a}+M_{2a})
\end{equation}
However, using \eqref{eq:sigma}, the total torque can be equivalently re-expressed as
\begin{equation}
    |\boldsymbol{\tau}_{tot}|=G(1-\sigma_{12})\frac{md}{|\boldsymbol{r}_M-\boldsymbol{d}|^2} (M_{1p}+M_{2p}) .
\end{equation}
This can be interpreted as Newton's constant $G$ now becoming material-dependent
\begin{equation}
 G_{12}\equiv  G(1-\sigma_{12})
\end{equation}
and the total torque taking the form 
\begin{equation}
    |\boldsymbol{\tau}_{tot}|=G_{12}\frac{md}{|\boldsymbol{r}_M-\boldsymbol{d}|^2} (M_{1p}+M_{2p})
\end{equation}
The corresponding angular acceleration of the torsion balance is now
\begin{equation}
    \ddot{\phi}=\frac{|\boldsymbol{\tau}_{tot}|}{I}=G_{12}(M_{1p}+M_{2p})\frac{1}{d|\boldsymbol{r}_M-\boldsymbol{d}|^2} 
\end{equation}
with $I=md^2$ the moment of inertia for the torsion. Precision measurements of distances, passive gravitational masses and angular acceleration $\ddot{\phi}$ lead to a numerical estimate for the material-dependent $G_{12}$, which in turn constrains $\sigma$.


\section{Appendix D: Overlap integral}
\label{app:overlap}
The overlap function between the Sodium and Cesium atoms is given by
\begin{equation}
    \mathcal{S}_{\text{Na-Cs}}=\int dV \Psi^{*}_{\text{Na}}\Psi_{\text{Cs}}
\end{equation}
where $\Psi_{\text{Na}},\Psi_{\text{Cs}}$ are the atomic wave-functions of Sodium and Cesium atoms respectively. For simplicity, we approximate these wave-functions as hydrogen-like orbitals $\Psi_{\text{Na}}\sim e^{-r/\alpha_{\text{Na}}}$  and $\Psi_{\text{Cs}}\sim e^{-r/\alpha_{\text{Cs}}}$ where $\alpha_{\text{Cs}}\approx 2.6 \r{A}$ and $\alpha_{\text{Na}}\approx 1.8 \r{A}$ are the Bohr radii for Cesium and Sodium respectively \cite{Slatter1964JChPh..41.3199S}. Solving the integral, we get 
\begin{equation}
  \mathcal{S}_{\text{Na-Cs}}\approx e^{-\frac{r}{\alpha_{\text{eff}}}}   
\end{equation}
with $\alpha_{\text{eff}}=\frac{\alpha_{\text{Na}}\alpha_{\text{Cs}}}{\alpha_{\text{Na}}+\alpha_{\text{Cs}}}\approx 1.06 \r{A}$. In order to ensure that the overlap is less than $15\%$, we get 
\begin{equation}
    r>2 \r{A}
\end{equation}
However, since our ultimate goal is to use lasers to create energy superposition states of different atomic orbitals, we require molecular binding energies on the order of $\mathcal{O}(10)$-$\text{GHz}$ to ensure that the molecule remains bound and do not disassociate. The binding energy $E_b$ of the $\text{Na-Cs}$ molecule can be computed as
\begin{equation}
    E_b= E_0 e^{-r/\alpha_{\text{eff}}}\Longrightarrow r\approx -\text{ln} \biggl(\frac{E_b}{E_0}\biggr)\alpha_{\text{eff}}
\end{equation}
where $E_0$ is an order of magnitude $\mathcal{O}(1\text{eV})$ constant, characteristic for this particular molecule. For binding energy $E_b$ corresponding to a frequency of $10 \text{GHz}$, we obtain
\begin{equation}
 r\approx -\text{ln} \biggl(\frac{4\cdot 10^{-5} \text{eV}}{\mathcal{O}(1)\text{eV}}\biggr)1.06 \r{A}\Longrightarrow    r\approx 10 \r{A}
\end{equation}
Thus, the optimal interatomic distance that satisfies both conditions is around $r\approx 10 \r{A}$. This distance ensures that the wave-function overlap is small enough while still allowing for a molecular binding energy in the GHz range.

\end{document}